\documentclass[aps,prd,superscriptaddress,10pt,nofootinbib,preprintnumbers]{revtex4} 
\usepackage{amsfonts,amssymb,amsmath}
\usepackage{graphicx}
\usepackage{hyperref}

\newcommand\DE[2]	{D^{E\,{#2}}_{\phantom{E\,}{#1}}}
\newcommand\DF[2]	{D^{F\,{#2}}_{\phantom{F\,}{#1}}}

\newcommand\be		{\begin{equation}}
\newcommand\ee		{\end{equation}}
\newcommand\corr[1]	{\langle{#1}\rangle}

\newcommand\req[1]	{(\ref{#1})}

\newcommand\p	{\partial}

\newcommand\cL	{\mathcal{L}}
\newcommand\cN	{\mathcal{N}}

\begin{document}

\preprint{YITP-18-11, IPMU18-0034}

\title{Revisiting Lorentz violation in Ho\v{r}ava gravity}

\author{Andrew Coates}
\affiliation{Theoretical Astrophysics, IAAT, University of T\"{u}bingen, Geschwister-Scholl-Platz, 72074, T\"{u}bingen, Germany}
\affiliation{School of Mathematical Sciences, University of Nottingham, University Park, Nottingham, NG7 2RD, United Kingdom}
\author{Charles Melby-Thompson}
\affiliation{Institut f\"{u}r Theoretische Physik und Astrophysik,
Julius-Maximilians-Universit\"{a}t W\"{u}rzburg,
Am Hubland, 97074 W\"{u}rzburg, Germany}
\affiliation{Department of Physics, Fudan University,
220 Handan Road, 200433 Shanghai, China}
\author{Shinji Mukohyama}
\affiliation{Center for Gravitational Physics, Yukawa Institute for Theoretical Physics, Kyoto University, 606-8502, Kyoto, Japan}
\affiliation{Kavli Institute for the Physics and Mathematics of the Universe (WPI), The University of Tokyo Institutes for Advanced Study, The University of Tokyo, Kashiwa, Chiba 277-8583, Japan}

\date{\today}

\begin{abstract}
In the context of Ho\v{r}ava gravity, the most promising known scenarios to recover Lorentz invariance at low energy are the possibilities that (1) the renormalization group flow of the system leads to emergent infrared Lorentz invariance, and (2) that supersymmetry protects infrared Lorentz invariance. A third scenario proposes that a classically Lorentz invariant matter sector with controlled quantum corrections may simply co-exist with Ho\v{r}ava gravity under certain conditions. 
However, for non-projectable Ho\v{r}ava gravity in $3+1$ dimensions it is known that, in the absence of additional structures, this mechanism is spoiled by unexpected power-law divergences. 
We confirm this same result in the projectable version of the theory by employing the recently found gauge-fixing term that renders the shift and graviton propagators regular. 
We show that the problem persists for all dimensions $D\geq 3$, and that the degree of fine tuning in squared sound speeds between a $U(1)$ gauge field and a scalar field increases with $D$. 
In particular, this difference in the zero external momentum limit is proportional to $\Lambda^{D-1}$ for $D\geq 3$, where $\Lambda$ is the ultraviolet momentum cutoff for loop integrals, while the power-law divergences are absent for $D=1$ and $D=2$. 
These results suggest that not only the gravity sector but also the matter sector should exhibit a transition to Lifshitz scaling above some scale, and that there should not be a large separation between the transition scales in the gravity and matter sectors. 
We close with a discussion of other more promising scenarios, including emergent Lorentz invariance from supersymmetry/strong dynamics, and pointing out challenges where they exist.
\end{abstract}

\maketitle

\section{Introduction}

Despite the empirical successes of General Relativity (GR), there are reasons to believe that it is incomplete.
The major stumbling block from a theoretical perspective is that quantum GR is not perturbatively renormalizable.
One approach to improving its UV properties is to relinquish Lorentz invariance at short distances~\cite{Horava:2009uw}, allowing the inclusion of higher order derivatives that render gravity renormalizable without violating unitarity. 
Renormalization group (RG) flow is organized around a fixed point with ``Lifshitz scaling'', which acts anisotropically on space and time, so that higher order spatial derivatives scale the same way as second order time derivatives in the ultraviolet~\cite{Horava:2009uw,Visser:2009fg}.%
\footnote{The Lifshitz scaling is also the origin of interesting cosmological implications of the theory, such as a novel mechanism for generating scale-invariant perturbations~\cite{Mukohyama:2009gg} and a solution to the flatness problem~\cite{Bramberger:2017tid}. See \cite{Mukohyama:2010xz} for a review of cosmology based on Ho\v{r}ava gravity.}
We refer to such theories as Ho\v{r}ava gravity.

Lorentz violations in the matter sector, however, are very tightly constrained. For example, the relative differences among speed limits for different species of matter fields are typically constrained to be smaller than $10^{-23}$ or so. Moreover, thanks to the recent multi-messenger observation of a binary neutron star merger, we now know that even the speed of gravitational waves cannot be different from that of photons by more than $10^{-15}$ or so. 
Therefore, naturalness demands a mechanism either protecting infrared Lorentz symmetry, or causing it to emerge~\cite{Kharuk:2015wga,Afshordi:2015smm} in the infrared.
If Ho\v{r}ava gravity or something similar is the correct description of gravity in the nature, those constraints require the RG flow of Lorentz-violating couplings to vanish rapidly in the IR~\cite{Chadha:1982qq}. 
This may be the case if supersymmetry protects infrared Lorentz invariance~\cite{GrootNibbelink:2004za}, and/or the system passes through a strongly coupled region which speeds up the otherwise logarithmic RG running of Lorentz-violating couplings toward zero~\cite{Anber:2011xf,Bednik:2013nxa}. 
We briefly discuss these possibilities in section~\ref{sec:solutions}. 

There is yet another possibility that, until recently, seemed quite promising: the mechanism proposed in~\cite{Pospelov:2010mp}.  
The basic statement of this mechanism is that, if one assumes that matter is relativistic at tree-level,
then Planck suppression of interactions between gravity and matter can lead
to a suppression of
Lorentz-violating corrections by the ratio
\begin{equation}
\frac{\delta c^2}{c^2} \sim 
\left(\frac{M_*}{M_p}\right)^2 \sim 10^{-18}\left( \frac{M_*}{10^{10}\, \mathrm{GeV}}\right)^2\,,
\end{equation}  
where $M_*$ is the scale suppressing the higher order spatial derivatives in the gravity sector. 
However, it was already shown in \cite{Pospelov:2010mp} that the vector sector of the non-projectable version of Ho\v{r}ava gravity does not benefit from this mechanism. Instead, one finds divergences of the form
\begin{equation}
\frac{\delta c^2}{c^2} \sim
\left(\frac{\Lambda}{M_p}\right)^2\,,
\end{equation}
where $\Lambda$ is the ultraviolet momentum cutoff. 
There, where the temporal gauge was used, this was attributed to the lack of Lifshitz scaling in the vector sector.%
\footnote{The authors state, however, that they verified their computations in a generalization of the $R_\xi$ gauge.}

The authors of~\cite{Pospelov:2010mp} proposed eliminating the power law divergences by introducing mixed derivatives into the Ho\v{r}ava action, in the form of terms quadratic in $\nabla_k K_{ij}$ ($K_{ij}$ is the extrinsic curvature of the spatial slices), see also \cite{Colombo:2015yha,Colombo:2014lta}.
In the non-projectable model, however, radiative corrections in the presence of mixed derivatives is expected to give rise to terms quadratic in time derivatives of the lapse $\cN$, leading to a new scalar degree of freedom that is unstable in the IR~\cite{Coates:2016zvg}. The viability of the mechanism, as presented in~\cite{Pospelov:2010mp} for the non-projectable theory, therefore needs a more in-depth analysis. We shall briefly discuss this point in section~\ref{sec:mixed derivatives}.

The present paper has two main goals. The first is to investigate the same mechanism in the projectable version of Ho\v{r}ava gravity in \(D+1\) dimensions. The projectable theory has technical advantages over the non-projectable theory, not the least of which is that it is known to be renormalizable~\cite{Barvinsky:2015kil}, while renormalizability of the non-projectable theory remains an open question.  A technical but crucial difference between the two theories is that only in the projectable case are all propagators in the gravity sector regular, 
explicitly respect the Lifshitz scaling in the ultraviolet, and have no instantaneous modes,%
\footnote{In the non-projectable theory there exist modes whose momentum space correlators diverge with respect to spatial momentum even at non-zero frequency $\omega$, which leads to instantaneous propagation.}
provided we adopt the gauge fixing of~\cite{Barvinsky:2015kil}. 
It is therefore beneficial to revisit Lorentz violation in the matter sector due to quantum corrections in the context of the projectable version of the theory. 
We shall confirm that the divergence structure is robust and that the problem persists in $D+1$ dimensions for any $D\geq 3$.

Our second goal is to discuss possible resolutions of these problems. 
These fall into three classes: mixed derivative terms in the projectable Ho\v{r}ava gravity, supersymmetry, and strong dynamics. The mixed derivative terms proposed in~\cite{Pospelov:2010mp} for the non-projectable model do not introduce a new scalar degree of freedom in the projectable theory, because there the offending term does not exist.
One is instead faced with infrared instabilities, imposing strong conditions on the value of $\lambda$.
The viability of the mechanism then reduces to the RG properties of $\lambda$, together with the existence of an analogue of the Vainshtein mechanism rendering the dynamics at $\lambda$ near 1 regular.
A second option --- first investigated in~\cite{GrootNibbelink:2004za} --- is the use of supersymmetry to control Lorentz violation in the infrared, and we comment on the possibility of combining the mechanisms of \cite{GrootNibbelink:2004za,Pospelov:2010mp,Pospelov:2013mpa} to suppress the loop contributions of the shift variable.
Finally, we briefly discuss the strong coupling mechanism studied in~\cite{Anber:2011xf,Bednik:2013nxa}.

The rest of the paper is organized as follows. 
Section~\ref{sec:setup} introduces our models, consisting of 
Ho\v{r}ava
gravity coupled to a free scalar and a $U(1)$ gauge field respectively, gives their propagators in the gauge-fixing of~\cite{Barvinsky:2015kil}, and lists the integrals that appear when calculating divergent contributions to the one-loop effective action.
Section~\ref{sec:loop corrections} computes the one-loop corrections to the scalar and gauge limiting speeds induced by the gravitational sector.
Possible solutions to the naturalness problem for Lorentz symmetry are discussed in section~\ref{sec:solutions}, and section~\ref{sec:summary} summarizes our conclusions and discusses interesting directions for future work.

\section{Setup and notation%
\label{sec:setup}}

Our treatment of Ho\v{r}ava gravity follows the setup of \cite{Barvinsky:2015kil} closely. 
We shall only take the extrinsic curvature terms and the terms which are of order $2D$ in spatial derivatives, as we are interested only in the UV contributions of gravitational degrees to loop integrals. 
We shall use the gauge-fixing prescription of \cite{Barvinsky:2015kil}, for which all propagators are regular. 
On the other hand, we needn't worry about introducing Faddeev-Popov ghost fields, since at one loop they do not contribute to the matter sector effective action. 
As we consider the projectable version of the theory, we are also free to set the lapse, $\cN$, to $1$ throughout the paper. 
We shall work in Euclidean signature, so that $\mathcal{L}_\mathcal{V}$ below has a minus sign relative to the real time action.
We then add a matter sector that (before Wick rotation) is Lorentz-invariant at the classical level, and investigate Lorentz-violating quantum corrections in the zero external momentum limit.

\subsection{Gravity action}

Ho\v{r}ava gravity starts with a fixed foliation of spacetime by constant time slices. In terms of a coordinate system $(t,x^i)$ respecting the foliation, it is built from the ADM fields $\cN$, $\cN_i$, and $g_{ij}$. 
We work solely with the so-called ``projectable'' theory, where $\cN=\cN(t)$  is constant on spatial slices. Its action is 
\be
S_{HL} = \frac{1}{2\kappa^2}\int dt\,d^{D\!}x\, \cN\!\sqrt{g}\Bigl[
	K_{ij}K^{ij} - \lambda K^2 + \cL_V + \cL_\text{g.f.}
\Bigr] \,,
\ee
where $K_{ij}=\frac{1}{2\cN}(\dot g_{ij}-\nabla_i \cN_j-\nabla_j \cN_i)$, and $\cL_V$ is built from the Riemann curvature tensor of the spatial slice using terms containing $2D$ spatial derivatives.
The coupling constants $\lambda$ and $\kappa^2$ are dimensionless under the UV scaling. 

To compute the leading divergent contributions to the matter effective action, it is enough to set the cosmological constant to zero and expand around a flat background:
\be
\cN = 1 \,,
\qquad\qquad
\cN_i = N_i \,,
\qquad\qquad
g_{ij} = \delta_{ij} + h_{ij} \,,
\ee
where $N_i$ and $h_{ij}$ are small perturbations. The gauge fixing term of~\cite{Barvinsky:2015kil} has two free parameters, $\sigma$ and $\xi$, and takes the form
\begin{equation}
\mathcal{L}_{\mathrm{g.f.}}
= \sigma F^i\mathcal{O}_{ij}F^{j}\,,
\end{equation}
where
\begin{eqnarray}
\mathcal{O}_{ij}&=&\Delta^{-D+2}\left[\delta_{ij}\Delta+\xi\partial_i\partial_j\right]^{-1},\\
F^i&=&\dot{N}^i+\frac{1}{2\sigma}\mathcal{O}^{-1}_{ij}\partial_kh_{jk}-\frac{\lambda}{2\sigma}\mathcal{O}^{-1}_{ij}\partial_jh\,.
\end{eqnarray}
This specific choice for the gauge fixing term decouples the shift and metric perturbations at quadratic order and renders all propagators regular. 
That this gauge-fixing term takes a universal form depending in a simple way on dimension can be traced to the fact that $N_i$ only appears in the kinetic term, which is independent of dimension and the choice of potential in the projectable theory.

\subsection{Matter action}
\label{subsec:matter}
Following~\cite{Pospelov:2010mp}, we take the matter sector to be Lorentz invariant at the HL scale, and evaluate the leading Lorentz-violating quantum corrections in the limit of small external momentum.
We will consider two types of matter: a massless scalar and a $U(1)$ gauge field, both taken to be Lorentz invariant in the UV.
After Wick rotation, the scalar action takes the form
\begin{equation}
 S_{\phi} = \frac{1}{2}\int d^{D+1}x\left[ (\partial_t\phi-N^i\partial_i\phi)^2 + \beta^2\delta^{ij}\partial_i\phi\partial_j\phi
+ O(h_{ij})
 \right]\,,
\label{eqn:Sphi}
\end{equation}
while the gauge field has the action
\begin{equation}
S_A = \int d^{D+1}x \left[
\frac{1}{2}(E_i - N^kF_{ki})^2 
+ \frac{\beta^2}{4}F_{ij}F_{ij} 
+ O(h_{ij}) \right] \,, \quad
E_i = \DE{i}{\mu}A_\mu\,, \quad 
 F_{ij} = \DF{ij}{\mu}A_\mu \,, 
\label{eqn:SA}
\end{equation}
with $\DE{i}{\mu}=\partial_0\delta^\mu_i-\partial_i\delta^\mu_0$ and $\DF{ij}{\mu}=\partial_i\delta^\mu_j-\partial_j\delta^\mu_i$. 
Here, we have suppressed the couplings to $h_{ij}$ because, as we argue in the following sections, $h$ loops contribute only logarithmic divergences.
The parameter $\beta$ ($>0$) is the bare propagation speed of $\phi$ and $A_{\mu}$, which we include for book-keeping purposes.

\subsection{Propagators}
Because of the particular form of the gauge fixing term, the shift propagator in Fourier space takes a simple form valid in any dimension:
\begin{equation} \label{eqn:shift-propagator}
\tilde{G}^{ij}_N(P_{\mu})
= a_1\left(\delta_{ij}-\hat{p}_i\hat{p}_j\right)\mathcal{P}_1(P_{\mu})+a_2 \hat{p}_i\hat{p}_j\mathcal{P}_2(P_{\mu})\,,
\qquad\qquad
  a_1=\frac{\kappa^2}{\sigma}\,,\quad
  a_2 = \frac{\kappa^2(1+\xi)}{\sigma}\,,
\end{equation}
where $P_{\mu}=(\omega,p_i)$ is the $4$-momentum, $\hat{p}_i=p_i/p$, $p=\sqrt{\delta^{ij}p_ip_j}$ and
\begin{equation}
\mathcal{P}_{1,2}(P_{\mu}) = 
  \frac{p^{2\left(D-1\right)}}{\omega^2 + \alpha_{1,2}^2 p^{2D}}\,,\qquad
\alpha_1^2=\frac{1}{2\sigma}\,,\quad
\alpha_2^2=\frac{(1-\lambda)(1+\xi)}{\sigma}\,.
\end{equation}
For the graviton $h_{ij}$,  any potential $\mathcal{L}_{\mathcal{V}}$ satisfying physically reasonable inequalities leads to a well-behaved Lifshitz dispersion relation, $\omega^2\propto k^{2D}$, for all modes. This leads to a Fourier-space propagator of the form
\begin{equation}
G^{h}_{ijk\ell}(P_\mu) = 
\sum_{n}M^{(n)}_{ijk\ell}\mathcal{P}'_{n}(P_{\mu})\,,
\label{eq:h propagator}
\end{equation}
where $M^{(n)}_{ijkl}$ runs over a set of tensor structures (projecting onto transverse traceless, divergenceless vector, and two scalar modes), and the $\mathcal{P}'_{n}$ are all of the form 
\begin{equation}
\mathcal{P}'_{n}(P_{\mu})=\frac{a'_{n}}{\omega^2+\alpha'^2_{n}p^{2D}} \,.
\end{equation}
The gauge-fixing term is such that the mixed propagator $\corr{N^ih_{jk}}$ vanishes.

For a scalar $\phi$ with the action (\ref{eqn:Sphi}), the Fourier space propagator is 
\begin{equation}
\tilde{G}_{\phi}(P_{\mu}) = \frac{1}{\omega^2+\beta^2p^2} \,,
\end{equation}
while a vector field $A_{\mu}$ in Feynman gauge with the action (\ref{eqn:SA}) has Fourier-space propagator 
\begin{equation}
\tilde{G}_{A\,\mu\nu}(P_{\rho}) = \frac{\eta^{\rm E}_{\mu\nu}}{\omega^2+\beta^2p^2}\,, \label{eqn:GAmunu}
\end{equation}
where $\eta^{\rm E}_{00}=\beta^2$, $\eta^{\rm E}_{0i}=\eta^{\rm E}_{i0}=0$ and $\eta^{\rm E}_{ij}=\delta_{ij}$.

\subsection{Loop integrals}
\label{subsec:integrals}

The leading divergence of the one-loop effective action for $\phi$ and $A_\mu$ will be derived in the next section. 
They can be expressed in terms of the following loop integrals:
\begin{equation}
\begin{gathered}
\begin{aligned}
 \mathcal{J}_1^{(I)} &\equiv a_I\int\frac{d^{D+1}P}{(2\pi)^{D+1}}\,\mathcal{P}_I(P_{\mu})\,, \qquad&
 \mathcal{J}_4^{(n)} &\equiv \int\frac{d^{D+1}P}{(2\pi)^{D+1}}\,\mathcal{P}'_n(P_{\mu})\,, \\
 \mathcal{J}_2^{(I)} &\equiv a_I\int\frac{d^{D+1}P}{(2\pi)^{D+1}}\,\omega^2\tilde{G}_{\phi}(P_{\mu})\mathcal{P}_I(P_{\mu})\,, &
 \mathcal{J}_5^{(n)} &\equiv \int\frac{d^{D+1}P}{(2\pi)^{D+1}}\,\omega^2\tilde{G}_{\phi}(P_{\mu})\mathcal{P}'_n(P_{\mu})\,, \\
 \mathcal{J}_3^{(I)} &\equiv a_I\int\frac{d^{D+1}P}{(2\pi)^{D+1}}\,p^2\tilde{G}_{\phi}(P_{\mu})\mathcal{P}_I(P_{\mu})\,, &
 \mathcal{J}_6^{(n)} &\equiv \int\frac{d^{D+1}P}{(2\pi)^{D+1}}\,p^2\tilde{G}_{\phi}(P_{\mu})\mathcal{P}'_n(P_{\mu})\,, 
\end{aligned}
\\
\mathcal{J}_7 \equiv \int\frac{d^{D+1}P}{(2\pi)^{D+1}}\,\tilde{G}_{\phi}(P_{\mu})\,. 
\end{gathered}
\end{equation}
These integrals are regularized and evaluated in appendix~\ref{app:integrals}.

\section{One-loop corrections in $D$ spatial dimensions%
\label{sec:loop corrections}}

Our goal is to compute the power-law divergent contributions to the shift in the limiting speeds of the fields $\phi$ and $A_\mu$.
This is done by computing the one-loop effective action for each field.
We will see in sections~\ref{subsec:graviton-scalar} and~\ref{subsec:graviton-gauge} that the divergence from $h_{ij}$ is logarithmic, and so the power-law divergence comes entirely from the shift $N_i$.
We therefore begin with loop corrections from the shift field.

\subsection{Shift field loop corrections to the scalar action}
\label{subsec:shift-scalar}

To compute the contribution of shift loops, we may set $h_{ij}=0$ and consider the action
\begin{equation}
S(\phi,N_i) = \frac{1}{2}\int d^{D+1}x\left[ (\partial_t\phi-N^i\partial_i\phi)^2 + \beta^2\delta^{ij}\partial_i\phi\partial_j\phi + N^i \mathcal{K}_{ij} N^j
\right]\,,
\end{equation}
where $\mathcal{K}_{ij}$ is the gauge-fixed kinetic term for $N^i$.
We are interested in the one-loop effective action for $\phi$, 
\begin{equation}
\Gamma_1(\phi) = \frac{1}{2}\mathrm{tr}\left(
	\log S^{(2)}[\phi] - \log S^{(2)}[0]\right)\,,
\end{equation}
where $S^{(2)}[\phi]$ is the fluctuation operator, with matrix components
\be
\bigl[S^{(2)}[\phi]\cdot f\bigr](x) = \int d^{D+1}y\,\frac{\delta^2S(\varphi)}{\delta\varphi(x)\delta\varphi(y)}\biggr\vert_{N_i=0} f(y) \,,
\ee
$\varphi$ stands for $(\phi,N_i)$, and $f$ is a test function.
Let $G$ denote the propagator matrix, and define the differential operator $U[\phi]$ by 
\begin{equation}\label{eq:Udef}
G\cdot S^{(2)}[\phi] = 1 + G\cdot U[\phi] \,.
\end{equation}
Note that $U[0] = 0$. Taylor expanding in $\phi$, we obtain
\begin{equation}
\Gamma_1(\phi) = \frac{1}{2}\mathrm{tr}\left(
	G\cdot U[\phi] - \frac{1}{2}G\cdot U[\phi]\cdot G\cdot U[\phi]
	+ O(\phi^3)\right)\,.
\label{eq:gamma 2}
\end{equation}
The components of $U$ are given by:
\begin{align}
\int d^{d+1}x\, d^{d+1}y\, f(x)\, U[\phi]_{\phi(x)\phi(y)} \, g(y) & = 0\,,\nonumber\\
\int d^{d+1}x\, d^{d+1}y\, f(x)\, U[\phi]_{\phi(x)N^i(y)} \, v^i(y) & =
  \int d^{d+1}x\, f(x) \Bigl[\partial_0\bigl(v^i(x)\partial_i\phi(x)\bigr)+\partial_i\bigl(v^i(x)\partial_0\phi(x)\bigr) \Bigr]\,,\nonumber\\
\int d^{d+1}x\, d^{d+1}y\, v^i(x)\, U[\phi]_{N^i(x)\phi(y)}\, f(y) & =
  \int d^{d+1}x\, f(x) \Bigl[\partial_0\bigl(v^i(x)\partial_i\phi(x)\bigr)+\partial_i\bigl(v^i(x)\partial_0\phi(x)\bigr) \Bigr]\,,\nonumber\\
\int d^{d+1}x\, d^{d+1}y\, v^i(x)\, U[\phi]_{N^i(x)N^j(y)}\, u^j(y) & =
  \int d^{d+1}x\, v^i(x)\,u^j(x)\, \partial_i\phi(x)\, \partial_j\phi(x) \,,
\end{align}
where $f$, $g$, $u^i$ and $v^i$ are test fields. 
Since the $\phi\,N_i$ Green's function vanishes, the effective action is given to quadratic order in $\phi$ by
\begin{equation}\label{QEA}
\Gamma_1^{(2)}(\phi) = 
\frac{1}{2}\mathrm{tr}\left( G^{ij}_N U_{N^iN^j} \right)
- \frac{1}{2}\mathrm{tr}\left( G_{\phi}U_{\phi N^i}G_N^{ij}U_{N^j\phi} \right) \,.
\end{equation}

Consider the first trace in \req{QEA}, corresponding to the diagram in figure \ref{Fig:1v}. 
In the Lifshitz gauge fixing \cite{Barvinsky:2015kil}, the (momentum space) propagator for $N^i$ is a sum of two terms as shown in (\ref{eqn:shift-propagator}). When $G_N^{ij}$ appears inside an integral multiplying a scalar expression, rotational invariance allows us to replace $p_ip_j\mapsto p^2\delta_{ij}/D$. The relevant integrals are defined in subsection~\ref{subsec:integrals} and evaluated in appendix~\ref{app:integrals}. The final form of the divergent contribution to the effective action from the first term of (\ref{QEA}) is therefore
\begin{equation}
\int d^{D+1}x\, G^{ij}_N(x,x) \partial_i\phi\partial_j\phi(x) =
\frac{1}{2}\int d^{D+1}x\, \zeta_3\delta^{ij}\partial_i\phi(x)\partial_j\phi(x)
\end{equation}
with
\begin{equation}
 \zeta_3 = \frac{\delta_{ij}}{D}G_N^{ij}(x,x) = \sum_{I=1}^2 A_I \mathcal{J}_1^{(I)}\,,
\qquad
A_1 = 1- \frac{1}{D},\quad A_2 = \frac{1}{D}\,.
\end{equation}

From the second term in (\ref{QEA}) (with diagram in figure \ref{Fig:2v}), we obtain a contribution to $\Gamma_1(\phi)$ of the form
\begin{equation}
-\frac{1}{2}
\int d^{D+1}x\int d^{D+1}y\left[
	(-\dot\phi\partial_j-\partial_j\phi\partial_\tau)^{(x)}G_\phi(x,y)
	(\dot\phi\partial_i+\partial_i\phi\partial_\tau+2\partial_i\dot\phi)^{(y)}G_N^{ij}(y,x)\right] \,. \label{mixed}
\end{equation}
From the shift symmetry of $\phi$ we know that all contributions to the Lagrangian must be of the form $D_1\phi D_2\phi$ where $D_1$ and $D_2$ are differential operators of order at least one. We are interested in the coefficients of $\dot\phi^2$ and $(\partial\phi)^2$. (Note that by time reversal, parity, and rotational symmetry, all other contributions to the effective action are less divergent, and have more derivatives acting on $\phi$). 
This leaves the leading divergent contribution
\begin{equation}
\frac{1}{2}\int d^{D+1}x\, 
\left[\zeta_1(\dot\phi)^2 + \zeta_2(\partial\phi)^2\right] \,.
\end{equation}
For $\zeta_1$ we find
\begin{equation}
\zeta_1 
= \left[\int d^{D+1}y \,\partial_j^{(x)}G_\phi(x,y)\partial_i^{(y)}G_N^{ij}(y,x)
\right]_\mathrm{div} = -\mathcal{J}^{(2)}_3 \,;
\end{equation}
note that $\mathcal{P}_1$ does not contribute because its prefactor $(\delta_{ij}-\hat p_i\hat p_j)$ inside $G_N^{ij}$ is transverse.
For $\zeta_2$, both $\mathcal{P}_1$ and $\mathcal{P}_2$ contribute as
\begin{equation}
\zeta_2
 = -\left[
     \frac{\delta_{ij}}{D}\int d^{D+1}y\partial^{(x)}_{\tau}G_{\phi}(x,y)\partial^{(y)}_{\tau}G_N^{ij}(y,x)\right]_\mathrm{div} 
= -\sum_{I=1}^2 A_I \mathcal{J}_2^{(I)}.
\end{equation}
We thus find the divergent contribution to the effective action quadratic 
both in $\phi$ and in derivatives to be
\begin{equation}
\Gamma_{1,\mathrm{div}}^{\mathrm{quad}}(\phi)
 = \frac{1}{2}\int d^{D+1}x \left[ \zeta_1\dot\phi^2
 + (\zeta_2+\zeta_3)(\partial\phi)^2 \right]
=
\frac{1}{2}\int d^{D+1}x\left[-\mathcal{J}_3^{(2)}\dot\phi^2
+ \sum_{I=1}^2 A_I(\mathcal{J}_1^{(I)}-\mathcal{J}_2^{(I)})(\partial\phi)^2\right]\,.
\end{equation}
Renormalizing so that $\phi$ has canonical kinetic term $\frac{1}{2}\dot\phi^2$, we may read off the $N^i$ loop contribution to the infrared shift in the limiting speed:
\begin{equation}
\delta c^2_{\phi} = \zeta_2+\zeta_3+\beta^2\delta Z_{\phi}^{(1)} = \zeta_2+\zeta_3-\beta^2\zeta_1
= \sum_{I=1}^2 A_I(\mathcal{J}_1^{(I)}-\mathcal{J}_2^{(I)}) + \beta^2\mathcal{J}_3^{(2)}
=\beta^2\left[\left(1-\frac{1}{D}\right)\mathcal{J}_{3}^{(1)}+\left(1+\frac{1}{D}\right)\mathcal{J}_{3}^{(2)}\right]\,, \label{eqn:deltac2phi}
\end{equation}
where $\delta Z_{\phi}^{(1)}$ is the $\mathcal{O}(\hbar)$ part of the field renormalization constant. The last equality follows from the first identity of equation~(\ref{eqn:identity-J}).

\subsection{Graviton loop corrections to scalar action}
\label{subsec:graviton-scalar}

\begin{figure}
  \centering
  \begin{minipage}[b]{0.4\textwidth}
  	\centering
    \includegraphics[scale=1.5]{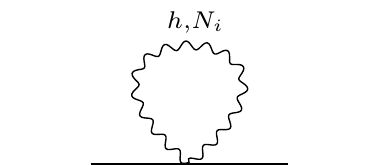}
    \caption{Single-vertex diagram.}\label{Fig:1v}
  \end{minipage}
  \begin{minipage}[b]{0.4\textwidth}
  	\centering
    \includegraphics[scale=1.5]{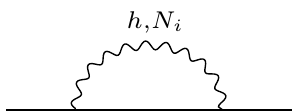}
    \caption{Two-vertex diagram.}\label{Fig:2v}
  \end{minipage}
  \hfill
\end{figure}

We now show that the one-loop contributions from the graviton $h_{ij}$ to $\delta c^2_{\phi}$ do not contain power-law divergences. 
One-loop gravitational contributions to the matter propagator come in two types, each of which is summed over the contributions of the two graviton modes. 
(By ``graviton'' we mean here the modes contributing to the $h_{ij}h_{k\ell}$ correlator \req{eq:h propagator}.)
The first is the single-vertex loop, and the second is the two-vertex loop describing the emission and subsequent reabsorption of a virtual graviton (see figures \ref{Fig:1v} and \ref{Fig:2v}).

In the first case, the loop integral involves only a factor of the graviton propagator, contributing a linear combination of integrals of the form $\mathcal{J}_4^{(n)}$, as defined in section~\ref{subsec:integrals}.
These integrals are shown in Appendix~\ref{app:integrals} to be free from power-law divergences. 

The second diagram involves in addition a factor of the scalar propagator. 
The $h\phi\phi$ vertices are a sum of $\dot\phi^2$ and $\p_i\phi\p_j\phi$ contributions, and as a result the diagrams are built from sums of the scalar and graviton propagators, weighted by the square of either the internal scalar energy or momentum. 
As a result, the leading divergence of these diagrams is given by a linear combination of integrals of the form $\mathcal{J}_5^{(n)}$ and $\mathcal{J}_6^{(n)}$, also defined in section~\ref{subsec:integrals}. 
Again, these contributions are free from power-law divergences, as shown in Appendix~\ref{app:integrals}. 
We have thus established that the power law divergence of the one-loop correction to the infrared limiting speed of the scalar field $\phi$ is given by (\ref{eqn:deltac2phi}).

\subsection{Shift field loop corrections to gauge field action}
\label{subsec:shift-gauge}

For shift loop contributions to the gauge field, we once again set $h_{ij}=0$, and consider
\begin{equation}
S = \int d^{D+1}x \left[
\frac{1}{2}(E_i - N^kF_{ki})^2 
+ \frac{\beta^2}{4}F_{ij}F_{ij}
+ N^i \mathcal{K}_{ij} N^j
		  \right]
+ S_\text{gauge-fixing}
\,,
\end{equation}
where $E_i$ and $F_{ij}$ are defined in (\ref{eqn:SA}), $\mathcal{K}_{ij}$ is the gauge-fixed kinetic term for $N^i$, 
and $\beta$ is the same bare propagation speed as before.
We adopt the Feynman gauge for $A$, so that the propagator is given by~\req{eqn:GAmunu}, although the final result is gauge invariant.

This time we are interested in the quadratic one-loop effective action for $A$:
\begin{equation}
\Gamma^{(2)}_1(A) =
\frac{1}{2}\mathrm{tr}\left( G\cdot U[A] - \frac{1}{2}G\cdot U[A]\cdot G\cdot U[A] \right)
 \,,
\end{equation}
where the $U$ operator has components
\begin{align}
 \int d^{d+1}x\, d^{d+1}y\, X_{\mu}(x)U[A]_{A_{\mu}(x)A_{\nu}(y)}Y_{\nu}(y) & =  0\,,\nonumber\\
 \int d^{d+1}x\, d^{d+1}y\, X_{\mu}(x)U[A]_{A_{\mu}(x)N^j(y)}v^j(y) & =
 \int d^{d+1}x\, X_{\mu}(x) \left[E_k \DF{jk}{\mu}v^{j} + F_{jk}\DE{k}{\mu}v^{j} + 2(\DE{k}{\mu}F_{jk})v^{j}\right]\,,\nonumber\\
 \int d^{d+1}x\, d^{d+1}y\, v^i(x)U[A]_{N^i(x)A_{\mu}(y)}Y^{\mu}(y) & =
 - \int d^{d+1}x\, v^i(x) \left[E_k \DF{ik}{\mu} X_{\mu}(x)+ F_{ik}\DE{k}{\mu} X_{\mu}(x)\right]\,,\nonumber\\
 \int d^{d+1}x\, d^{d+1}y\, v^i(x)U[A]_{N^i(x)N^j(y)}u^j(y) & =
  \int d^{d+1}x\, v^i(x)u^j(x)F_{ik}F_{jk}\,.
\end{align}
Here, $\DE{i}{\mu}=\partial_0\delta^\mu_i-\partial_i\delta^\mu_0$ and $\DF{ij}{\mu}=\partial_i\delta^\mu_j-\partial_j\delta^\mu_i$ as in subsection~\ref{subsec:matter}, and $X_{\mu}$, $Y_{\mu}$, $u^i$ and $v^i$ are test fields. 
We first compute:
\begin{equation}
\frac{1}{2}\mathrm{tr}\left(G\cdot U[A]\right) 
= \frac{1}{2}\mathrm{tr}_N\left(G_N\cdot U_{NN}[A]\right)
= \frac{1}{2}\int d^{D+1}x\, G^{ij}_N(x,x)\, (F_{ik}F_{jk})(x)
= \frac{1}{4}\zeta_3^A\int d^{D+1}x\, F_{ij}F_{ij}\,,
\end{equation}
where $\zeta_3^A=2\zeta_3^\phi$.
Next, we have the $E_i^2$ contribution. 
Pulling this out of the analog of (\ref{mixed}), we have
\begin{equation}
-\left.\frac{1}{2}\mathrm{tr}\left(G_N\cdot U_{NA}\cdot G_{A}\cdot U_{AN}\right)\right|_{E^2}
= \frac{1}{2}\int d^{D+1}x\int d^{D+1}y\, 
\left(E_{\ell}\DF{j\ell}{\mu}\right)^{(x)}G^A_{\mu\nu}(x,y)
\left(E_k \DF{ik}{\nu}\right)^{(y)}G_N^{ij}(y,x) \,.
\end{equation}
As before, the dominant divergence is
\begin{equation}
\frac{1}{2} \int d^{D+1}x\, E_k E_\ell Z^{k\ell}\,,
\end{equation}
where $Z^{k\ell}=(\tilde{Z}^{k\ell}+\tilde{Z}^{\ell k})/2$ and 
\begin{eqnarray}
\tilde{Z}^{k\ell} &= \int d^{D+1}y\, \DF{j\ell}{\mu(x)}G^A_{\mu\nu}(x,y)
\, \DF{ik}{\nu(y)}G_N^{ij}(y,x) .
\end{eqnarray}
By rotational invariance, $Z^{k\ell} = \delta_{k\ell}\zeta_1^A$, where
\begin{align}
\zeta_1^A = \frac{1}{D}\delta_{k\ell}Z^{k\ell}
&= \frac{1}{D}\int d^{D+1}y\, \int d^{D+1}y\, \DF{jk}{\mu(x)}G^A_{\mu\nu}(x,y)
\, \DF{ik}{\nu(y)}G_N^{ij}(y,x) \nonumber\\
&= -\frac{1}{D}\int\frac{d\omega\,d^Dp}{(2\pi)^{D+1}}
\frac{(p_j\eta^{\rm E}_{k\nu}-p_k\eta^{\rm E}_{j\nu})(p_i\delta_k^\nu-p_k\delta^\nu_i)}
{\omega^2+\beta^2p^2}\left[a_1(\delta_{ij}-\hat{p}_i\hat{p}_j)\mathcal{P}_1+a_2\, \hat{p}_i\hat{p}_j\mathcal{P}_2\right] \nonumber\\
&= -\left(1-\frac{1}{D}\right)\int\frac{d\omega\,d^Dp}{(2\pi)^{D+1}}
\frac{p^2}{\omega^2+\beta^2p^2}\,
\delta^{ij}\left[a_1(\delta_{ij}-\hat{p}_i\hat{p}_j)\mathcal{P}_1+a_2\, \hat{p}_i\hat{p}_j\mathcal{P}_2\right] \nonumber\\
&= -\left(1-\frac{1}{D}\right)\sum_{I=1}^2 \mathcal{J}_3^{(I)}\,.
\end{align}
Similarly, the $F^2$ term is
\begin{equation}
\frac{1}{2}\int d^{D+1}x\int d^{D+1}y\,
\bigl(F_{j\ell}\DE{\ell}{\mu}\bigr)^{(x)}G^A_{\mu\nu}(x,y)
\left(F_{ik}\DE{k}{\nu}\right)^{(y)}G_N^{ij}(y,x) \,,
\end{equation}
with leading divergence
\begin{equation}
\frac{1}{4}\int d^3x\, F_{ik}F_{j\ell}Z^{ik,j\ell} \,,
\end{equation}
where $Z^{ik,j\ell}=[(\tilde{Z}^{ik,j\ell}-\tilde{Z}^{ik,\ell j}-\tilde{Z}^{ki,j\ell}+\tilde{Z}^{ki,\ell j})+(\tilde{Z}^{j\ell,ik}-\tilde{Z}^{j\ell,ki}-\tilde{Z}^{\ell j,ik}+\tilde{Z}^{\ell j,ki})]/8$ and 
\begin{equation}
\tilde{Z}^{ik,j\ell} = 2\left.\int d^{D+1}y\, \DE{\ell}{\mu(x)}G^A_{\mu\nu}(x,y)
\DE{k}{\nu(y)}G_N^{ij}(y,x) \right|_{\mathrm{div}} \,.
\end{equation}
Using $\eta^{\rm E}_{00}=\beta^2$, $\eta^{\rm E}_{0i}=\eta^{\rm E}_{i0}=0$, and $\eta^{\rm E}_{ij}=\delta_{ij}$, we obtain
\begin{eqnarray}
\tilde{Z}^{ik,j\ell} &=& 
-2\int\frac{d\omega\,d^Dp}{(2\pi)^{D+1}}
\frac{(\omega\delta^\mu_\ell-\delta^\mu_0p_\ell)\eta^{\rm E}_{\mu\nu}}
{\omega^2+\beta^2p^2}
(\omega\delta_k^\nu - p_k\delta_0^\nu)\left[a_1\left(\delta_{ij}-\hat{p}_i\hat{p}_j\right)\mathcal{P}_1+a_2 \hat{p}_i\hat{p}_j\mathcal{P}_2\right]  \nonumber\\
&=&
-2\int\frac{d\omega\,d^Dp}{(2\pi)^{D+1}}
\frac{\omega^2\delta_{k\ell}+\beta^2p_kp_\ell}{\omega^2+\beta^2p^2}\left[a_1\left(\delta_{ij}-\hat{p}_i\hat{p}_j\right)\mathcal{P}_1+a_2 \hat{p}_i\hat{p}_j\mathcal{P}_2\right]  \,.
\end{eqnarray}
For each tensor structure inside the integral we can replace
\begin{equation}
p_ip_j\rightarrow \frac{1}{D}p^2\delta_{ij} \,,
\qquad\qquad
p_ip_jp_kp_\ell\rightarrow 0 \,,
\end{equation}
where the latter comes from contracting with $F_{ik}$.
Taking into account index symmetries, the final result can be written as
$Z^{ik,j\ell}=\zeta_2^A(\delta_{ij}\delta_{k\ell}-\delta_{i\ell}\delta_{kj})/2$. The contribution to $\zeta_2^A$ from the $\mathcal{P}_1$ term is 
\begin{equation}
-2
\int\frac{d\omega\,d^{D}p}{(2\pi)^{D+1}}
\frac{p^{2(D-1)}(\omega^2\left(1-\frac{1}{D}\right) + \beta^2p^2\left(\frac{1}{D}\right))}
{(\omega^2+\alpha_1^2p^{2D})(\omega^2+\beta^2p^2)}
= -2\left[\left(1-\frac{1}{D}\right)\mathcal{J}_1^{(1)}-\left(1-\frac{2}{D}\right)\beta^2\mathcal{J}_3^{(1)}\right]
\end{equation}
while the contribution from the $\mathcal{P}_2$ term is
\begin{equation}
-\frac{2}{D}
\int\frac{d\omega\,d^Dp}{(2\pi)^{D+1}}
\frac{\omega^2p^{2(D-1)}}
{(\omega^2+\alpha_1^2p^{2D})(\omega^2+\beta^2p^2)}
= -\frac{2}{D}\mathcal{J}_2^{(2)} \,.
\end{equation}
This gives the net result
\begin{eqnarray}
\zeta_2^A &=-2\left[\left(1-\frac{1}{D}\right)\mathcal{J}_1^{(1)}-\beta^2\left(1-\frac{2}{D}\right)\mathcal{J}_3^{(1)}+\frac{1}{D}\mathcal{J}_2^{(2)}\right]
\,.
\end{eqnarray}
Combining these results and the contribution from field renormalization, one obtains
\begin{eqnarray}
 \delta c^2_A
  & = & \left(\zeta_2^A+\zeta_3^A\right)-\beta^2\zeta_1^A
  = \frac{2}{D}\left(\mathcal{J}_1^{(2)}-\mathcal{J}_2^{(2)}\right)
  + \beta^2\left[2\left(1-\frac{2}{D}\right)\mathcal{J}_3^{(1)}+\left(1-\frac{1}{D}\right)\sum_{I=1}^2\mathcal{J}_3^{(I)}\right]\nonumber\\
 & = & \beta^2\left[\left(3-\frac{5}{D}\right)\mathcal{J}_3^{(1)}+\left(1+\frac{1}{D}\right)\mathcal{J}_3^{(2)}\right]\,. \label{eqn:deltac2A}
\end{eqnarray} 
Once again, the last equality follows from the first identity of (\ref{eqn:identity-J}).

\subsection{Graviton loop corrections to gauge field action}
\label{subsec:graviton-gauge}

The argument given in section~\ref{subsec:graviton-scalar} that the graviton modes do not contribute power law divergences to $\delta c^2_\phi$ carries over to the gauge field without modification.
Therefore the power law divergence of the one-loop correction to the infrared limiting speed of the gauge field $A_{\mu}$ is given by (\ref{eqn:deltac2A}).

\subsection{Lorentz-violating quantum corrections}
\label{sec:power law LV}

To summarize, we find that the shift loop contributions (\ref{eqn:deltac2phi}) and (\ref{eqn:deltac2A}) generate power law divergences in the difference in squared sound speeds between a $U(1)$ gauge field and a scalar field, of the form 
\begin{equation}
\delta c^2_A - \delta c^2_\phi=\frac{2\left(D-2\right)}{D}\mathcal{J}_3^{(1)} \,.
\end{equation}
This vanishes only for $D=1$, where $\mathcal{J}_3^{(1)}$ vanishes because of the absence of the first term ($\propto \left(p^2\delta_{ij}-p_ip_j\right)=0$) in (\ref{eqn:shift-propagator}), and for $D=2$.
As shown in Appendix~\ref{app:integrals}, the integral $\mathcal{J}_3^{(1)}$ diverges as $\sim \Lambda^{D-1}$, where $\Lambda$ is the ultraviolet momentum cutoff. Therefore, we have a naturalness problem for all $D\geq 3$. 
For $D=3$ (where $M_p^2 = 1/\kappa^2$), this reduces to
\begin{equation}
\delta c^2_A - \delta c^2_\phi=\frac{1}{4\pi^2}\left(\frac{\Lambda}{M_p}\right)^2 
\,.
\end{equation}

\section{Possible solutions%
\label{sec:solutions}}

In this section we discuss some possible resolutions coming from modification of the gravity theory. Since we have shown that the naturalness problem for Lorentz-violating couplings only becomes worse for $D>3$, in what follows we shall focus on $D=3$.

\subsection{Mixed derivatives}
\label{sec:mixed derivatives}

The first proposed solution to this naturalness problem, presented in the original paper, was to introduce mixed derivative terms~\cite{Pospelov:2010mp}.
In non-projectable Ho\v{r}ava gravity this means allowing terms quadratic in the following objects: 
\begin{equation}\label{eq:mixedderivativeingredients}
 \nabla_{i}K_{jk}\,, \quad
  r_{ij} \equiv \frac{1}{\mathcal{N}}
  \left(\dot{R}_{ij}-\mathcal{N}^k\nabla_kR_{ij}-R_{ik}\nabla_j\mathcal{N}^k-R_{kj}\nabla_i\mathcal{N}^k\right)\,, \quad
  \mathcal{A}_i \equiv \frac{1}{2\mathcal{N}}
  \left(\dot{a}_i-\mathcal{N}^j\nabla_ja_i-a_j\nabla_i\mathcal{N}^j\right)\,,
\end{equation}
where $a_i\equiv \partial_i\ln\mathcal{N}$. Unfortunately, the term quadratic in the last of these,  
\begin{equation}\label{eq:sigma1def}
\frac{M_p^2 \sigma_1}{M^2}g^{ij}\mathcal{A}_i\mathcal{A}_j \,,
\end{equation}
generates a new scalar degree of freedom possessing an IR instability \cite{Coates:2016zvg}. If we preserve the stability criteria of the theory without mixed derivatives while avoiding ghosts, the instability comes from the relative sign of the \(\mathcal{A}^2\) term and the term proportional to \(a^2\). 

In the projectable version of Ho\v{r}ava gravity, on the other hand, the lapse function $\mathcal{N}$ is constant in space and thus the offending term \req{eq:sigma1def} does not arise. 
However, the projectable version of Ho\v{r}ava gravity without mixed derivative terms is also known to have an infrared gradient instability. Since mixed derivative terms are irrelevant in the infrared, their presence does not affect the nature of infrared instabilities. For the theory to be phenomenologically viable, the time scale of the infrared instability must be longer than either the Hubble time scale or the standard Jeans time scale. In $3+1$ dimensions, this leads to the following non-trivial condition on the RG flow of $\lambda$~\cite{Mukohyama:2010xz}:
\begin{equation}
 0 < \frac{\lambda-1}{3\lambda-1} < \max\left[\frac{H^2}{k^2}\,,|\Phi|\right]\quad
  \mbox{for}\ H<k<\min\left[M\,,\frac{1}{0.01 mm}\right]\,, \label{eqn:condRGflow}
\end{equation}
where $k$ is the momentum scale of interest, $H$ is the Hubble expansion rate of the background, $\Phi$ is the Newton potential and $M$ is the Lifshitz scale. This in particular implies that $\lambda$ should run towards $1$ sufficiently rapidly in the infrared. In the limit $\lambda\to 1$ the perturbative expansion breaks down, and a fully nonlinear analysis is needed. In general such nonlinear analyses are technically difficult. Fortunately, in some simple cases, such as stationary spherically symmetric configurations~\cite{Mukohyama:2010xz} and nonlinear superhorizon perturbations~\cite{Izumi:2011eh,Gumrukcuoglu:2011ef}, fully nonlinear analyses have been performed, and it has been shown that general relativity (plus ``dark matter as integration constant''~\cite{Mukohyama:2009mz,Mukohyama:2009tp}) is recovered in the limit $\lambda\to 1$. It is therefore worthwhile investigating the naturalness problem in the projectable theory with mixed derivative terms. 

A similar constraint can be placed in the non-projectable theory, giving
\begin{equation}
 0< M^2 \frac{\alpha}{\sigma_1} < \frac{1}{4} \max\left[H^2, k^2 |\Phi|\right]\quad
\mbox{for}\ H<k<\min\left[M\,,\frac{1}{0.01 mm}\right] \,, 
  \label{eqn:condRGflow-mixed}
\end{equation}
where \(\sigma_1\) was defined in \eqref{eq:sigma1def} and \(\alpha\) is the coefficient of \(a_i a^i\) \cite{Blas:2009yd}. 
Treating this is as a constraint on the IR behavior of $\alpha$, one obtains
\begin{equation}
\alpha \lesssim 10^{-104} \left(\frac{10^{10}\mathrm{GeV}}{M}\right)^2\left(\frac{\sigma_1}{1}\right)\max\left[H^2/H_0^2,k^2|\Phi|/H_0^2\right]\,, 
\end{equation}
where $H_0\sim 10^{-42}\mathrm{GeV}$ is the present value of $H$. 

If one likes to avoid technical complications due to nonlinearity of the scalar graviton in establishing the low-energy recovery of general relativity, then the $U(1)$ extension~\cite{Horava:2010zj} may provide a way out since there is no scalar graviton in the $U(1)$ extension with projectability.\footnote{On the other hand, there is a scalar graviton in the non-projectable version of the $U(1)$ extension~\cite{Lin:2013tua,Mukohyama:2015gia}.} 
In this case, matter can couple universally to a gauge-invariant variable constructed from the gauge field and the so-called Newtonian prepotential at low energy, e.g. after integrating out heavy fields.\footnote{See Appendix~C of \cite{Lin:2013tua} for a simple example.} 
It has been shown that all the solar system tests carried out so far are satisfied in a large region of the parameter space~\cite{Lin:2013tua}. It is therefore worthwhile investigating the naturalness problem in the context of the projectable $U(1)$ extension, with or without mixed derivative terms, in future work.

\subsection{Supersymmetry}
\label{subsec:susy}
One promising mechanism to protect infrared Lorentz invariance is to impose supersymmetry~\cite{GrootNibbelink:2004za}. 
While the supersymmetry (SUSY) algebra is usually treated as an extension of Poincar\'e symmetry, supersymmetry does not imply Lorentz symmetry. 
It was shown in~\cite{GrootNibbelink:2004za,Bolokhov:2005cj} that, in a 4d theory with $\cN=1$ SUSY, if all fields are charged under a gauge symmetry and there is at most one $U(1)$ gauge field, then Lorentz violating (LV) operators have dimension $\ge 5$.
If CPT is imposed,%
\footnote{The CPT theorem assumes Lorentz invariance. Therefore, in a Lorentz-violating theory CPT invariance is a non-trivial condition we are free to impose.}
the lower bound on LV operator dimensions becomes $6$.
We assume CPT in what follows since CPT violation is rather strongly constrained, e.g., by observation of polarized gamma rays from distant gamma-ray bursts (see \cite{Toma:2012xa} and references therein). 

If SUSY is a UV symmetry of nature then it must be broken at some point, generating LV effects below the breaking scale. 
These effects must be suppressed in order to maintain IR Lorentz invariance. 
When SUSY is softly broken at scale $m_\text{soft}$ dimension~$4$ operators are generated, but are suppressed by the ratio $m^2_\mathrm{soft}/\Lambda_\text{LV}^2$,
where $\Lambda_\text{LV}$ is the LV scale~\cite{Pospelov:2010mp,Pospelov:2013mpa}. 
This protects IR Lorentz symmetry provided that $m_\mathrm{soft}\ll \Lambda_\text{LV}$, and that $\Lambda_\text{LV}$ is sufficiently large that the dimension-$6$ operators satisfy observational constraints. 
Taking energy dependence of gamma-ray burst light curves~\cite{AmelinoCamelia:1997gz}, the suppression scale of the dimension $6$ operator, which is expected to be of order $\Lambda_\text{LV}$ unless fine-tuned, is constrained as $M_{\rm QG,2}>10^{11}{\rm GeV}$~\cite{Ackermann:2009aa}. 
Unfortunately, as it stands this argument only works if the UV theory is supersymmetric, which requires Ho\v{r}ava gravity to be supersymmetric as well.
It is not known whether such a supersymmetrization exists.

An alternative was suggested in~\cite{Pospelov:2013mpa}, where it was proposed that supersymmetry breaking in the matter sector could be initiated by LV gravity.
As a result, both SUSY breaking and Lorentz violation arise from the same coupling, and the contributions of graviton loops are controlled by a single dimensionless ratio,
\be
\frac{\delta c}{c}
\sim \frac{\Lambda_\text{HL}^2}{M_p^2} \,,
\qquad
m^2_\text{soft} \sim \Lambda_\text{UV}^2\cdot\frac{\Lambda_\text{HL}^2}{M_p^2} 
\,.
\ee
The analysis of~\cite{Pospelov:2013mpa} does not, however, take into account the loop contribution from the vector modes of the gravity sector. 
As these were the source of difficulty in our analysis and in the analysis of~\cite{Pospelov:2010mp}, an alternative mechanism --- such as the mixed derivative terms discussed in section~\ref{sec:mixed derivatives} --- is presumably required to guarantee naturalness.

We comment briefly on an additional possibility. 
The power-law divergence of $\delta c^2$ discussed in section~\ref{sec:power law LV} can be attributed, in part, not to a non-Lifshitz dispersion relation of $N_i$, but rather to the scaling behavior of the matter field.
The divergence can, however, be reduced by increasing the dimension of the inverse scalar propagator, corresponding to the introduction of higher dimension LV operators.
The authors of~\cite{Pospelov:2010mp} avoided introducing higher dimension LV operators because in typical field theories, loop effects lead to $O(1)$ values of $\delta c^2$.
Since SUSY protects the infrared value of $\delta c^2$, however, if the split in scales between SUSY breaking and Lorentz violation is sufficiently large, it may be possible to suppress LV effects from matter loops and gravity loops simultaneously.
If SUSY is broken as in~\cite{Pospelov:2013mpa} by interaction with Ho\v{r}ava gravity, it is possible that naturalness can be preserved without mixed derivative terms.
We leave a detailed investigation of this scenario to future work.

\subsection{Strong dynamics}
\label{subsec:strongdynamics}

Another proposal is that if one allows for strong dynamics then Lorentz invariance can be restored in the infra-red \cite{Bednik:2013nxa,Anber:2011xf,Chadha:1982qq}. Suppose we have a matter sector with modes which have different sound speeds and are coupled with coupling constant $g$. 
Parametrizing their relative speed difference as
\begin{equation}
\eta=\frac{\beta_1}{\beta_2}-1,
\end{equation}
near $\beta_1\approx\beta_2\approx c$ one has the typical behavior
\begin{equation}
\frac{\partial \eta}{\partial \log\mu}=\frac{b}{4\pi^2 c^3} g^2 \eta \,,
\end{equation}
where $\mu$ is the renormalization scale and $b=O(1)$~\cite{Anber:2011xf}. 
If $g$ is a classically marginal coupling then it will run logarithmically, $g^2\sim(A+B\log M/\mu)^{-1}$, causing $\eta$ to run according to 
\begin{equation}
\eta\sim \frac{1}{(A+B\log M/\mu)^{\#}} \,,
\end{equation}
where $M$ is some large mass scale~\cite{Anber:2011xf}.
In this case $\eta$ may flow to zero, but much too slowly to be consistent with observational constraints.
If the coupling constant does not run, however, then $\eta$ runs with a power law whose exponent is proportional to $g^2$, as
\begin{equation}
\eta\propto \left(\frac{\mu}{M}\right)^{\frac{b}{4\pi^2c^3}g^2}\,.
\end{equation}
Thus, when two modes spend a long RG time near a strongly coupled fixed point, their sound speeds rapidly converge as we flow to the infrared \cite{Bednik:2013nxa,Anber:2011xf}.%
\footnote{Another construction that speeds up the flow is to introduce many additional matter/gauge fields~\cite{Anber:2011xf}.
This can lead effectively to a power-law running, but it requires the addition of a large hidden sector, together with a mechanism decoupling it at experimental energies from the Standard Model fields. 
It is therefore unlikely that this possibility is feasible phenomenologically.}

Finally, if a similar mechanism functions for the gravity sector as well, it might help fulfill the phenomenological constraints on the renormalization group flow presented in (\ref{eqn:condRGflow}) and (\ref{eqn:condRGflow-mixed}).

\bigskip

Before concluding it is worth commenting on the possibility of reducing the degree of divergence by introducing additional fields unrelated to any symmetry (unlike the \(U(1)\) extension or a hypothetical supersymmetric version of Ho\v{r}ava gravity). 
As it is the vector modes which are problematic the symmetry of the background requires we introduce a vector field $B_i$. 
To modify the propagators, we should couple $B_i$ to $N_i$ locally (the theory should be local before gauge fixing), but such a coupling is strongly constrained by the foliation preserving diffeomorphisms and CPT symmetry.
Given the existence of a suitable coupling there is another problem. In order to suppress the problematic divergence requires that the diagonalised vector system have propagators that scale like \(1/k^{2 z}\) (or be suppressed even further) in the UV.
This then requires that the new spin-1 mode have a Lifshitz scaling propagator with \(z\geq D\). Without some symmetry principle one could then write infinitely many independent terms, and thus the theory would be non-predictive and hence we don't find this a viable option.

\section{Summary%
\label{sec:summary}}

This paper considered the mechanism proposed in \cite{Pospelov:2010mp} for obtaining a matter sector with natural IR Lorentz invariance coupled to non-projectable Ho\v{r}ava gravity, and applied it to the projectable theory.
We showed that, as in the non-projectable model, in the absence of additional ingredients this mechanism is spoiled by contributions from the vector modes. 
Our analysis, valid for any number of spatial dimensions $D$, shows that the problem exists for any $D>2$, and grows worse as $D$ increases.%
\footnote{Curiously, although power counting suggests a linear dependence on $\Lambda$ for $D=2$, these terms cancel we find only logarithmic divergences in this case. 
We have only verified this special feature for spins 0 and 1.}
In particular, the naturalness problem cannot be avoided by compactifying a higher-dimensional theory.
Our computation furthermore used the gauge-fixing of \cite{Barvinsky:2015kil}, in which all propagators are regular and have Lifshitz scaling, further supporting the results of \cite{Pospelov:2010mp}, and suggesting that something more than the irregularity of the shift propagator lies at the heart of the mechanism's failure. Namely, the source of difficulty is simply the lack of Lifshitz scaling in the matter sector in the UV, combined with the fact that the shift propagator in the UV is less suppressed than the graviton propagator.

We discussed several possible approaches to this problem. 
One modification proposed already in \cite{Pospelov:2010mp} was the addition of mixed derivative terms, which for the non-projectable theory was found in \cite{Coates:2016zvg} to generate a new propagating degree of freedom with IR instabilities.
As a result, the success of the mechanism relies on whether the IR couplings satisfy the stability condition \req{eqn:condRGflow-mixed}. 
The offending extra mode does not exist in the projectable model with mixed derivatives, but the gradient (IR) instability of the original model without mixed derivatives persists. 
For this model to have viable phenomenology requires both an analogue of the Vainshtein mechanism, and for the couplings to flow in the IR to values satisfying the phenomenological constraints (\ref{eqn:condRGflow-mixed})~\cite{Mukohyama:2010xz}. 
We leave the investigation of these issues to future work.

A second possibility, proposed in~\cite{Pospelov:2010mp,Pospelov:2013mpa,GrootNibbelink:2004za}, is based on supersymmetry.
It is therefore interesting to seek a supersymmetric version of Ho\v{r}ava gravity. 
(For work relevant to this direction, see \cite{Redigolo:2011bv,Pujolas:2011sk}  on the supersymmetrization of field theories with Lifshitz scaling and the \ae{}ther vector, respectively.) 
Meanwhile, one conceivable solution would combine UV supersymmetry in the matter sector (without Lorentz invariance), together with soft SUSY breaking by interaction with Ho\v{r}ava gravity.

The remaining mechanism appearing in the literature posits that RG flow is responsible for IR emergent Lorentz symmetry \cite{Bednik:2013nxa,Anber:2011xf,Chadha:1982qq}. 
For weakly-coupled models the flow is too slow to be phenomenologically viable, requiring that the RG flow pass near a strongly-coupled fixed point.
Such behavior is expected in some beyond-standard-model scenarios, such as walking technicolor.
Detailed models with the necessary properties thus need to be developed and studied. 

\smallskip
Let us now comment on the scenario of emergent Lorentz symmetry and its relation to our results. In the literature~\cite{Chadha:1982qq,Anber:2011xf,Bednik:2013nxa}, emergent Lorentz symmetry was investigated without including higher spatial derivative terms. We thus need to extend the analysis to systems with higher spatial derivative terms, so that all fields enjoy the Lifshitz scaling with a common value of the dynamical critical exponent $z$ in the UV. That is, we need to consider dispersion relations satisfying $\omega^2\simeq k^{2z}/M^{2(z-1)}$ for $k\gg M$ and $\omega^2\simeq c_s^2k^2$ for $k\ll M$, where $M$ is the suppression scale of the higher spatial derivative term. In principle, $M$ can be different for different species. The main result of the present paper, namely the failure of the mechanism proposed in \cite{Pospelov:2010mp}, clearly excludes the case with $M_{\rm matter}\gg M_{\rm grav}$ since in this case the matter sector would have the $z=1$ scaling all the way up to $\Lambda\sim M_{\rm matter}$ as in the mechanism of \cite{Pospelov:2010mp}. On the other hand, if $M_{\rm matter}\sim M_{\rm grav}$ then the main result of the present paper does not apply to the scenario of emergent Lorentz symmetry and it is generically expected that Lorentz invariance in the matter sector may emerge at low energy as a consequence of the RG flow. In summary, the results of the present paper suggest that, not only the gravity sector, but also the matter sector should exhibit Lifshitz scaling above some scale, and that there should not be a large separation between the transition scales in the gravity and matter sectors. 

Our discussion so far has been limited to the matter sector. The recent multi-messenger observations of a binary neutron star merger~\cite{GBM:2017lvd}, however, constrain the difference between the speed of gravitational waves $c_{\rm gw}$ and the speed of light $c_{\gamma}$ to the level of $O(10^{-15})$~\cite{Creminelli:2017sry,Ezquiaga:2017ekz,Baker:2017hug}. In the context of the scenario of emergent Lorentz symmetry, it is therefore important to see under what conditions (if any) there can be a fast RG flow toward $c_\text{gw}=c_\gamma$.

It is, finally, worth comparing naturalness of Lorentz symmetry to some better-known naturalness problems in cosmology: the cosmological constant and curvature problems. 
The cosmological constant is relevant in the IR, so the cosmological constant problem should be more severe than the curvature problem (as evidenced by the ability of inflation to solve the latter and not the former). 
Since $\lambda$ and $c_\text{gw}$ are classically marginal in the IR, power-counting suggests that the naturalness problems discussed in this paper will also be less severe than the cosmological constant problem.

\begin{acknowledgments}
The authors would like to thank M. Pospelov for helpful correspondence.
SM thanks the LMPT for hospitality. 
AC is grateful for the hospitality of the Yukawa Institute of Theoretical Physics where this work started and also to the University of Nottingham. 
The work of AC was undertaken in part as an overseas researcher under a Short-Term Fellowship of the Japan Society for the Promotion of Sciences. 
The work of CMT was supported by the Thousand Young Talents Program, Fudan University, and a Humboldt Research Fellowship from the Alexander von Humboldt Foundation. The work of SM was supported by Japan Society for the Promotion of Science (JSPS) Grants-in-Aid for Scientific Research (KAKENHI) No. 17H02890, No. 17H06359, and by World Premier International Research Center Initiative (WPI), MEXT, Japan.  
\end{acknowledgments}

\appendix

\section{Regulator and loop integrals} \label{app:integrals}

We regulate the integrals introduced in section~\ref{subsec:integrals} as follows. We first integrate over all $\omega=P_0$, and then cut off the upper limit of the $p$ integrals at $p=\Lambda$, where $p=\sqrt{\delta^{ij}p_ip_j}$ and $p_i=P_i$. While this regularization scheme is not gauge-invariant and modifies the long-range behavior of propagators, it suffices for the purpose of computing the general behavior of ultraviolet divergences, as we are taking $\Lambda^D>>\Lambda M_*^{D-1}$, where $M_*$ is the scale at which anisotropic scaling begins to dominate. With this regularization scheme, we have
\begin{eqnarray}
a^{-1}\mathcal{J}_1 &= &\int\frac{d\omega\,d^Dp}{(2\pi)^{D+1}}\frac{p^{2(D-1)}}{\omega^2+\alpha^2p^{2D}}
= \int\frac{d^Dp}{2^{D+1}\pi^D}\frac{p^{D-2}}{\alpha} \nonumber \\
&= &\frac{1}{2^{D}\pi^{D/2}\Gamma\left(D/2\right)\left(D-1\right)}\cdot \frac{\Lambda^{2\left(D-1\right)}}{2\alpha}\,, \\
a^{-1}\mathcal{J}_2 &= &\int\frac{d\omega\,d^Dp}{(2\pi)^{D+1}}
\frac{\omega^2p^{2(D-1)}}{(\omega^2+\beta^2p^2)(\omega^2+\alpha^2p^{2D})}
=  \int\frac{d^Dp}{2^{D+1}\pi^{D}}\frac{p^{2(D-1)}}{\alpha p^D+\beta k}\nonumber \\
&= &\frac{1}{2^{D}\pi^{D/2}\Gamma\left(D/2\right)\left(D-1\right)}\cdot\left(\frac{\Lambda^{2(D-1)}}{2\alpha}-\Lambda^{(D-1)}\frac{\beta}{\alpha^2}+\frac{\beta^2}{\alpha^3}\log{\frac{\alpha}{\beta}\Lambda^{D-1}}\right) +\mathcal{O}(1/\Lambda^{D-1})\,, \\
a^{-1}\mathcal{J}_3
&=&\int\frac{d\omega\,d^Dp}{(2\pi)^{D+1}}
\frac{p^{2D}}{(\omega^2+\beta^2p^2)(\omega^2+\alpha^2p^{2D})}
= \int \frac{d^Dp}{2^{D+1}\pi^{D}} \frac{p^{D-1}}{\alpha\beta(\alpha p^{D}+ k \beta)} \nonumber \\
&=& \frac{1}{2^{D}\pi^{D/2}\Gamma\left(D/2\right)\left(D-1\right)}\cdot\left(\frac{\Lambda^{D-1}}{\alpha^2\beta}-\frac{1}{\alpha^3}\log{\left(\frac{\alpha}{\beta}\Lambda^{D-1}\right)}\right)+\mathcal{O}(1/\Lambda^{D-1})\,, 
\end{eqnarray}
where $\mathcal{J}_{1,2,3}$, $a$ and $\alpha$ stand for $\mathcal{J}_{1,2,3}^{(I)}$, $a_I$ and $\alpha_I$ ($>0$) ($I=1,2$), respectively, 
\begin{eqnarray}
{a'}^{-1}\mathcal{J}_4
&= &\int\frac{d\omega\,d^Dp}{(2\pi)^{D+1}}\frac{1}{\omega^2+{\alpha'}^2p^{2D}} 
= \int \frac{d^Dp}{2^{D+1}\pi^{D}}\frac{1}{{\alpha'} p^D} \nonumber \\
&= &\frac{1}{2^{D}\pi^{D/2}\Gamma\left(D/2\right){\alpha'}}\log\left(\frac{\Lambda}{m}\right)\,, \\
{a'}^{-1}\mathcal{J}_5 
&= & \int\frac{d\omega\,d^Dp}{(2\pi)^{D+1}}\frac{1}{\omega^2+{\alpha'}^2p^{2D}}\frac{\omega^2}{\omega^2+\beta^2 p^2}
= \int \frac{d^Dp}{2^{D+1}\pi^{D}}\frac{1}{{\alpha'} p^D+\beta k}\nonumber \\
&= &\frac{1}{2^{D}\pi^{D/2}\Gamma\left(D/2\right)\left(D-1\right){\alpha'}}\log\left(\frac{{\alpha'}}{\beta}\Lambda^{D-1}\right)+\mathcal{O}(1/\Lambda^{D-1})\,, \\
{a'}^{-1}\mathcal{J}_6
&= &\int\frac{d\omega\,d^Dp}{(2\pi)^{D+1}}\frac{1}{\omega^2+{\alpha'}^2p^{2D}}\frac{p^2}{\omega^2+\beta^2 p^2}
= \int \frac{d^Dp}{2^{D+1}\pi^{D}}\frac{1}{{\alpha'}\beta p^{D-1}}\frac{1}{{\alpha'} p^D+\beta k}\nonumber \\
&=& -\frac{1}{2^{D}\pi^{D/2}\Gamma\left(D/2\right){\alpha'}\beta^2\left(D-1\right)}\log\left(\frac{{\alpha'}}{\beta}m^{D-1}\right)+\mathcal{O}(m^{D-1},1/\Lambda^{D-1}) 
 = (\textrm{finite in $\Lambda$})\,, 
\end{eqnarray}
where $\mathcal{J}_{4,5,6}$, $a'$ and $\alpha'$ stand for $\mathcal{J}_{4,5,6}^{(n)}$, $a'_n$ and $\alpha'_n$ ($>0$), respectively, and
\begin{equation}
 \mathcal{J}_7
= \int\frac{d\omega\,d^Dp}{(2\pi)^{D+1}}
\frac{1}{\omega^2+\beta^2p^2} = \frac{1}{2^{D}\pi^{D/2}\Gamma\left(D/2\right)\left(D-1\right)}\frac{\Lambda^{D-1}}{\beta}\,.
\end{equation}
The integrals $\mathcal{J}_4$ and $\mathcal{J}_6$ suffer from infrared divergences, so we have introduced an IR cutoff $m$. As integral $\mathcal{J}_6$ has only infrared divergences its inclusion is not required for an analysis of UV properties, but presumably it is required if one wants to keep track of the cancellation of infrared divergences.

We also have the following identities
\begin{equation}
\mathcal{J}_2 + \beta^2\mathcal{J}_3 = \mathcal{J}_1\,,
\quad
\mathcal{J}_5 + \beta^2\mathcal{J}_6 = \mathcal{J}_4 \,,
\label{eqn:identity-J}
\end{equation}
and the following approximate relation that holds up to parts which vanish as $\Lambda \to \infty$ and $m\to 0$.
\begin{equation}
 {a'}^{-1}\mathcal{J}_5 + \alpha^2a^{-1}\mathcal{J}_3 = \mathcal{J}_7
  + \frac{1}{2^D\pi^{D/2}\Gamma(D/2)(D-1)}\left(\frac{1}{\alpha'}-\frac{1}{\alpha}\right)
  \left[\ln\left(\frac{\alpha}{\beta}\Lambda^{D-1}\right)+\mathcal{O}(\Lambda^{-(D-1)},m^{D-1})\right]\,.
\end{equation}

\end{document}